# A Versatile Post-Doping Towards Two-Dimensional Semiconductors

Yuya Murai[1], Shaochun Zhang[1], Takato Hotta[1], Zheng Liu[2], Yasumitsu Miyata[3], Toshifumi Irisawa[4], Yanlin Gao[5], Mina Maruyama[5], Susumu Okada[5], Hiroyuki Mogi[5], Tomohiro Sato[5], Shoji Yoshida[5], Hidemi Shigekawa[5], Takashi Taniguchi[6], Kenji Watanabe[7], and Ryo Kitaura[1,*]

[1]*Department of Chemistry, Nagoya University, Nagoya, Aichi 464-8602, Japan*

[2]*Innovative Functional Materials Research Institute, National Institute of Advanced Industrial Science and Technology (AIST), Nagoya, Aichi 463-8560, Japan*

[3]*Department of Physics, Tokyo Metropolitan University, Hachioji, Tokyo 192-0397, Japan*

[4]*Device Technology Research Institute, National Institute of Advanced Industrial Science and Technology (AIST), Tsukuba, Ibaraki 305-8568, Japan*

[5]*Department of Physics, Graduate School of Pure and Applied Sciences, University of Tsukuba, 1-1-1 Tennodai, Tsukuba, Tsukuba 305-8571, Japan*

[6]*International Center for Materials Nanoarchitectonics, National Institute for Materials Science, 1-1 Namiki, Tsukuba 305-0044, Japan*

[7]*Research Center for Functional Materials, National Institute for Materials Science, 1-1 Namiki, Tsukuba 305-0044, Japan*

Corresponding Author: R. Kitaura, r.kitaura@nagoya-u.jp

**Abstract**

We have developed a simple and straightforward way to realize controlled post-doping towards 2D transition metal dichalcogenides (TMDs). The key idea is to use low-kinetic energy dopant beams and a high-flux chalcogen beam at the same time, leading to substitutional doping with controlled dopant densities. Atomic-resolution transmission electron microscopy has revealed that dopant atoms injected toward TMDs are incorporated substitutionally into the hexagonal framework of TMDs. Electronic properties of doped TMDs (Nb-doped $WSe_2$) have shown drastic change, p-type action with more than two orders of magnitude increase in on current. Position-selective doping has also been demonstrated by the post-doping toward TMDs with a patterned mask on the surface. The post-doping method developed in this work can be a versatile tool for 2D-based next-generation electronics in the future.

**INTRODUCTION**

The current limitation in Si-based devices, arising from the adverse short-channel effect, has led to searching for new materials for the next-generation electronic devices. The short-channel effect becomes serious when the electric field originating from the source/drain electrodes is comparable to that of the gate electrode. In this case, the gate electric field cannot control the channel current properly, and adverse effects, such as a decrease of threshold voltage and degradation of subthreshold characteristics, inevitably emerge. Various approaches, including double-gate field-effect transistors (FET) and FinFET, are now being investigated to overcoming the short-channel effect.(*1, 2*) Among the various approaches, one of the promising ways is to use new material, particularly atomically-thin semiconductors (two-dimensional (2D) semiconductors), as semiconductor channels.(*3*)

The discovery of 2D semiconductors has opened a golden avenue toward next-generation nanoelectronics.(*4, 5*) Research on 2D materials started with gapless graphene, and graphene research soon led to the isolation of various 2D semiconductors, such as transition metal dichalcogenides (TMDs, for example, $MoS_2$, $WSe_2$, etc.), whose bandgap ranges in energy from 1~2 eV.(*6, 7*) 2D semiconductors possess a uniform thickness of less than 1 nanometer without, ideally, any dangling bonds at the surface, and the ultra-thin uniform structure of 2D semiconductors is ideal to realize ultra-short channel devices. Past researches on 2D semiconductors have already demonstrated that 2D semiconductors work as high on/off ratio FETs with carrier mobility up to the order of $10^2$ cm$^2$/Vs at room temperature.(*8-10*) In conjunction with the recent advancement of growth techniques and the successful formation of ohmic contacts, 2D semiconductors provide an excellent opportunity to develop next-generation nanoelectronics.(*11, 12*)

The accurate post-doping method for 2D materials is crucial for the development of 2D-material-based electronic devices. The success in Si devices is based on reliable post-doping techniques; p/n-doped Si works as electrodes/channels in Si-based metal-oxide-semiconductor FETs (MOSFETs) in integrated circuits. In the fabrication of Si MOSFETs, p/n dopants are incorporated through ion implantation or diffusion of dopants, forming electrodes and channels at desired locations. However, the same technique cannot be applied for doping into 2D semiconductors because high-energy ions or high-temperature diffusion processes can damage the ultra-thin structure of 2D semiconductors. Doping toward 2D materials has, therefore, been performed through mixing dopants during growth processes.(*13, 14*) Although this is a versatile way to dope p/n dopants substitutionally to frameworks of 2D materials, the development of accurate post-doping is essential to realize position-controlled doping for future 2D-semiconductor-based devices.

In this work, we have developed a straightforward way to realize controlled post-doping towards 2D transition metal dichalcogenides (TMDs), including $MoSe_2$, $WSe_2$, etc. Figure 1 shows a schematic representation of the doping process developed in this work. The key idea is to use dopant beams (Nb, Re, etc.) with low kinetic energy generated by thermal evaporation of high-melting-point metals. Kinetic energy distribution $P(E_{kin}, \beta)$ of atoms in dopant beams is expected to be proportional to $\exp(-\beta E_{kin})$, where $E_{kin}$ and $\beta$ represent kinetic energy and inverse temperature ($1/k_B T$, $k_B$: Boltzmann constant, $T$: temperature), respectively. In high-melting-point metals, such as Nb and Re, we need a temperature around the melting point (2750 and 3458 K for Nb and Re) to have beam flux high enough for doping processes. In this case, the distribution of $E_{kin}$ ranges typically from several hundreds of meV to 1~2 eV, leading to substitutional doping without significantly damaging the original framework of host 2D semiconductors. Also, we supply a Se beam with a high flux rate (Se/Nb > 3000) during the whole doping process to heal Se vacancies and enhance structural reconstruction to form well-regulated substitutionally doped TMDs. In this method, accurate and position-selective doping becomes possible through accurate control of the dopant beam flux with a mask layer patterned on 2D TMDs, leading to 2D-based next-generation electronics in the future.

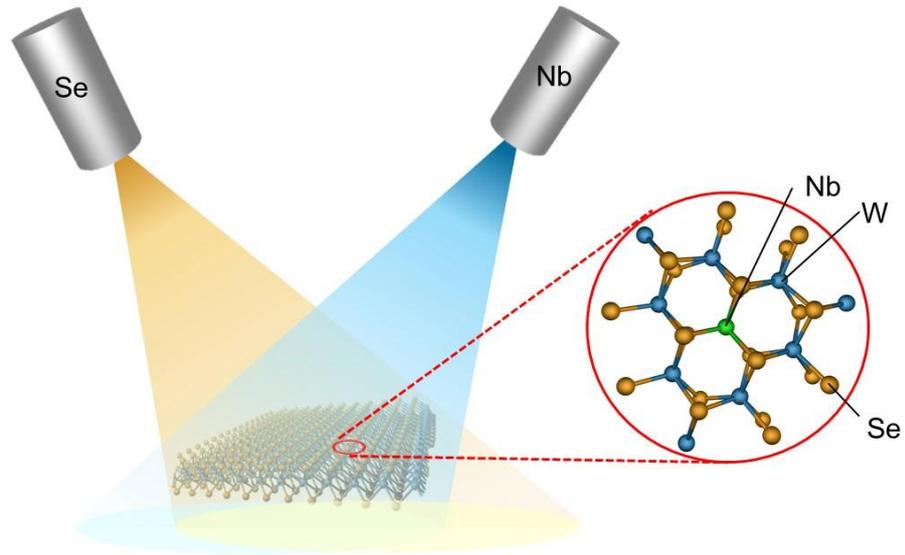

Figure 1. A schematic representation of the post-doping process. A pre-deposited crystal of TMD (in this case, $WSe_2$) is irradiated with a dopant (Nb) and a chalcogen (Se) beam, which leads to substitutional post-doping of Nb toward $WSe_2$.

## RESULTS

**Structural characterization of Nb-doped WSe$_2$**

Figures 2a and b show atomic-resolution high-angle annular dark-field (HAADF)-scanning transmission electron microscope (STEM) images of a monolayer WSe$_2$ after 3- and 6-minute exposures of Niobium (Nb) beams with a flux of 0.67 monolayer/hour. Although optical images show almost no change after the doping process (Fig. S1), HAADF-STEM images show significant change. Spot-like contrasts observed in the images can be roughly categorized into three groups: the brightest, the middle, and the darkest ones. The brightest and the darkest ones correspond to W and single Se atoms, respectively; single Se atoms probably arise from the formation of Se defects induced by electron-beam irradiation. The middle contrasts observed at the W positions correspond to Nb atoms substituted into the framework of WSe$_2$. In HAADF-STEM images, image contrasts strongly depend on the atomic number (Z-contrast), and substitutions of W by Nb yield darker contrasts than those of W, as seen in the HAADF-STEM images. The line profile along the dotted line in the image well matches that of the simulated image (Fig. 2c), which further confirms the substitution of W by Nb.(*15*) We carefully checked HAADF-STEM images and found that no W or Nb atoms exist on top of 2D layers; the additional contrast from surface adsorbed W/Nb atoms has not been observed in any HAADF-STEM images. Energy-dispersive X-ray spectroscopy has revealed that the Nb substitution occurs uniformly over the WSe$_2$ samples (Fig. S2). Substitution of Se by Nb is energetically unfavorable, and therefore, Nb atoms should predominantly occupy W sites in the current doping process. Direct counting the number of Nb atoms substituted gives doping rates of 3.4 and 7.4 % for 3- and 6-minute exposures, respectively. The corresponding doping rate, in this case, is 1.2 %/min. Because the determined doping rate is slow enough, it is possible to control the degree of doping precisely, and we can easily make the doping rate slower just by using smaller Nb flux in this method.

Whereas most Nb atoms are isolated to form one-atom-sized dark contrasts in Figs. 2a and b, some Nb atoms form clusters in WSe$_2$ (particularly in Fig. 2a). To see if Nb substituted in WSe$_2$ favors the formation of clusters or not, we calculated alloying degree $J$ defined as follows.(*16*)

$$J_W = P_{obs} / P_{rand} \times 100 \%$$

where $P_{obs}$ and $P_{rand}$ represent the ratio of the averaged coordination number of Nb to the total coordination number (in this case, 6) and the atomic ratio of Nb in the examined region. Whereas $J_W$ of 100 % means no preference in neighboring atoms for Nb substituted in WSe$_2$, $J_W$ smaller/larger than 100 % means Nb atoms prefer isolated/clustering. Counting 300 Nb atoms yields $J_W$ of 97 and 92 % for doping rates of 3.4 and 7.4 %, respectively, suggesting that the present Nb substitution occurs almost randomly. The observed random doping indicates that Nb atoms may exist as single atoms or tiny clusters composed of 2~3 atoms in Nb beams, and after landing on a WSe$_2$ layer, Nb atoms are immediately incorporated into the WSe$_2$ layer without significant surface diffusion to form large Nb clusters.

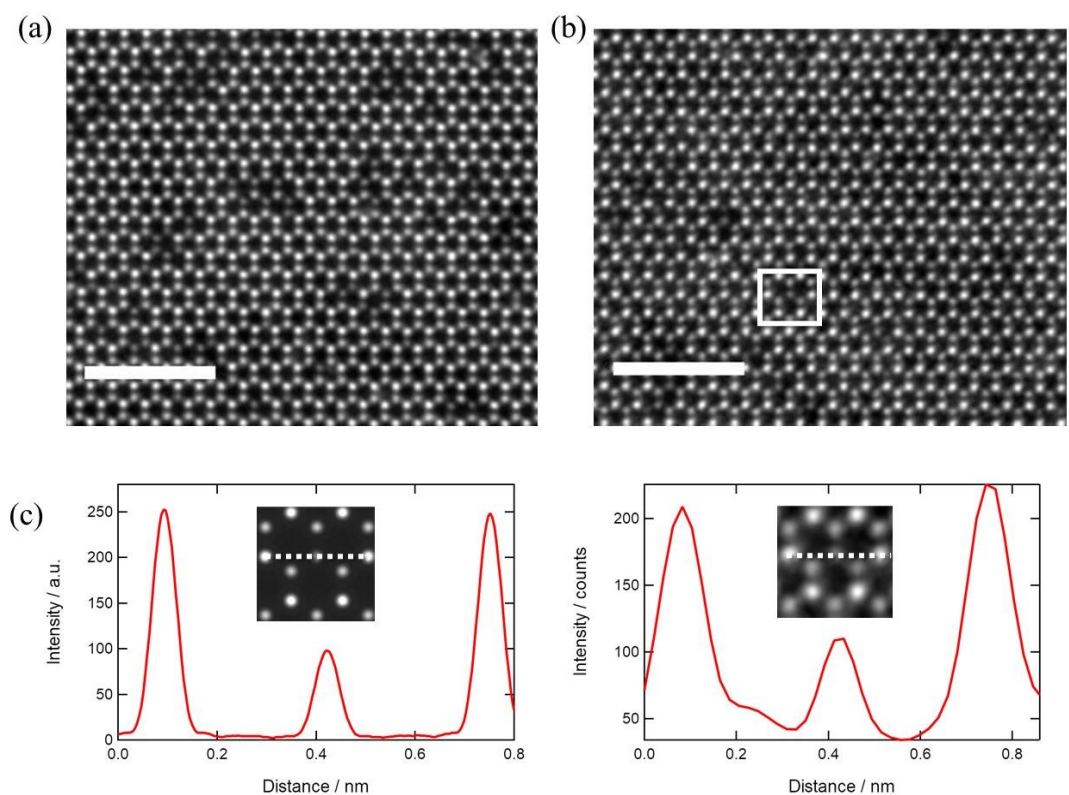

Figure 2 (a) and (b) a typical HAADF-STEM image of monolayer WSe$_2$ after 3 and 6 minutes doping of Nb. After applying subtract background with rolling ball radius of 2 nm, bandpass filter was applied to filter out large and small structures, whose size is 2 and 0.06 nm, respectively. All these image processings were performed with ImageJ.(*20*) The scale bars correspond to 2 nm. (c) a simulated (left) and an observed (right) HAADF-STEM image and the corresponding line profiles. Line profiles were made along the dotted lines shown in the images. The observed image used to make the line profile is a cutout image from the white box in the image (b).

Figure 3a shows room-temperature photoluminescence (PL) spectra of monolayer WSe$_2$ before and after a 6-minute Nb beam exposure. As clearly demonstrated, Nb substitutional doping causes a significant decrease in PL intensity; the PL peak arises from radiative recombination of bright excitons at K/-K valley. The observed reduction in PL intensity is consistent with carrier doping induced by the substitutional doping because increased carrier results in an increase in non-radiative recombination through the Auger process.(*17, 18*) One important thing here is that surface of WSe$_2$ is clean without any increase in surface roughness after the doping, and the quenching in PL is not caused by impurity-induced unintentional doping or growth of metallic NbSe$_2$ on the surface. In addition to the decrease in PL intensity, PL peaks shift to the red side, which is consistent with the previous works on substitutional Nb doping; the increase in hole density and possible strain induced by Nb doping can cause the PL redshift. Corresponding Raman spectra (Fig. 3b), where Raman peaks arising from A'$_1$ and 2LA are seen, show intensity reduction, being also compatible with the earlier works.(*17, 19*) Both TEM-based structural characterizations and optical responses shown above clearly demonstrate that injected Nb atoms substitutionally replace W atoms in WSe$_2$. Typical AFM image before and after doping (Fig. 3c) shows that the formation of second layers does not occur.

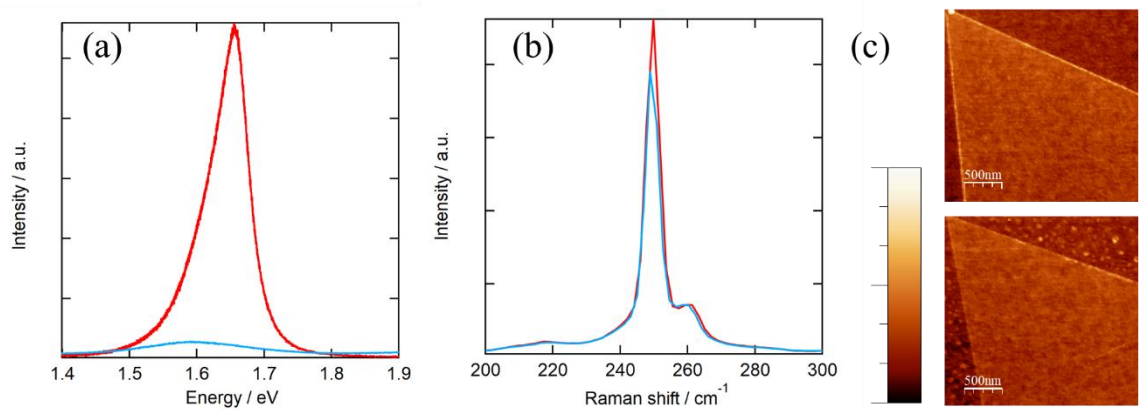

Figure 3 (a), (b), and (c) a typical PL, Raman spectra, and AFM images of a pristine monolayer WSe$_2$ and the monolayer WSe$_2$ after 6-minutes doping of Nb. Red and blue curves correspond to spectra of pristine sample and Nb-doped sample, respectively. We used the excitation wavelength of 532 nm to measure the PL and Raman spectra. Intensities of the Raman spectra are normalized by the Si peak at 520 cm$^{-1}$. The maximum and minimum heights shown by the color scale in (c) are 4.5 and 0 nm, respectively.

**Mechanism of the doping process**

To address the mechanism of the present doping process, we have performed an ab-initio molecular dynamics (MD) simulation, where an Nb atom with a kinetic energy of about 300 meV is injected towards a monolayer WSe$_2$ (similar results were obtained with different kinetic energies, Fig. S3). Figure 4 shows snapshots of the structural change upon the injection (a movie is available as supplemental movie files). As seen in the snapshots, the Nb atom hits a W atom, followed by the release of the W atom from the original position. The W atom released moves around the original position during simulation time (3 ps). As mentioned above, we observed no W or Nb atom on the WSe$_2$ layer after doping processes, which means that released W atoms are incorporated into WSe$_2$ again. We perform the doping at 823 K with the supply of Se with a high flux rate of 2 Å/sec (Se/Nb > 3000), and structural reconstruction to recover the original hexagonal network occurs under this condition. The high-flux Se supply during the doping processes is probably essential to form the hexagonal network structure because injecting Nb atoms alone results in the formation of not the hexagonal network but mirror twin boundaries.(*21*) As seen in the ab-initio MD calculation, Nb atoms injected are rapidly incorporated into WSe$_2$ without significant surface diffusion. In this case, the formation of Nb clusters is not likely, which is consistent with the analyses of STEM images, random Nb doping.

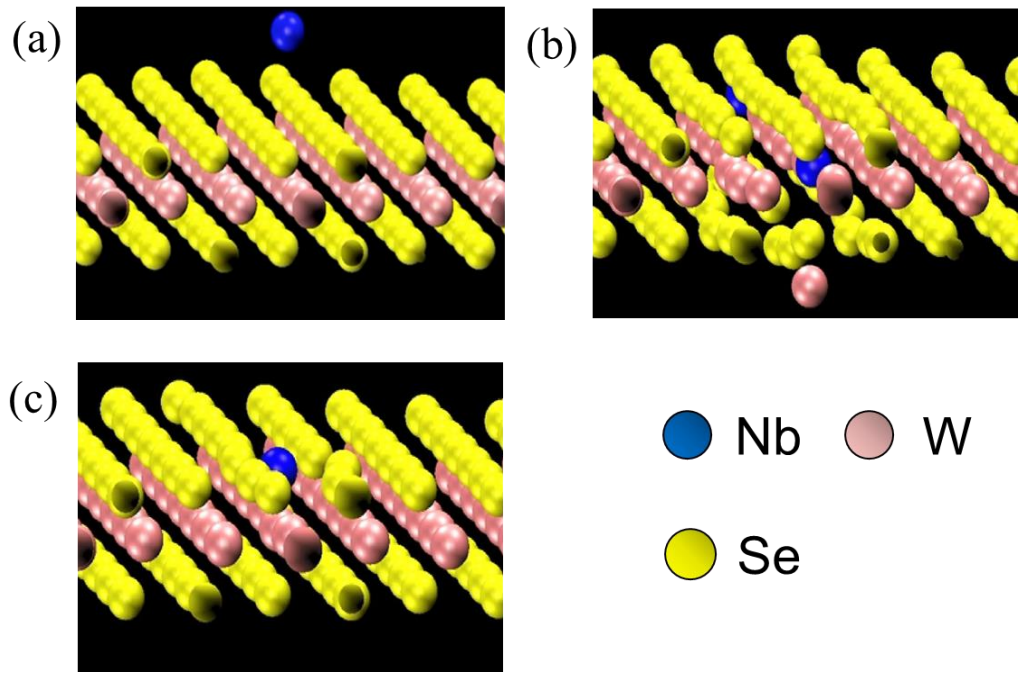

Figure 4 Snapshots of ab-initio molecular dynamics simulation of Nb atom injected toward monolayer WSe$_2$. (a), (b), and (c) show a snapshot after 0, 40, and 100 fs, respectively.

One of the advantages of the present doping method is position-controlled doping. For this purpose, we deposit a patterned mask on a monolayer WSe$_2$ by electron beam lithography with an inorganic resist (Hydrogen Silsesquioxane, HSQ); HSQ is stable up to a process temperature of 823 K. Figure 5 shows optical and Raman images of WSe$_2$ with HSQ pattern after 6-minutes doping. As clearly seen, the region exposed to the dopant beam shows quenching of Raman intensity, giving dark contrasts in the Raman image. It should be noted that the edges of patterns in the Raman images are sharp, which means that the substitutional doping occurs only at the exposed region. This is because the doping processes are performed at 823 K, which is low enough to inhibit the thermal diffusion of doped Nb atoms in the framework of WSe$_2$. Given that lateral TMD heterostructures grown at high temperatures (~ 1000 K) give sharp junction interfaces, the temperature needed for thermal diffusion within the framework is supposed to be much higher than the temperature in the current doping processes.(*22, 23*) The high-temperature required for the thermal diffusion means that substitutional doping with nanometer-scale precision is possible in the current method, leading to the possibility of realizing arrays of FETs with ultra-short 2D channels in the future.

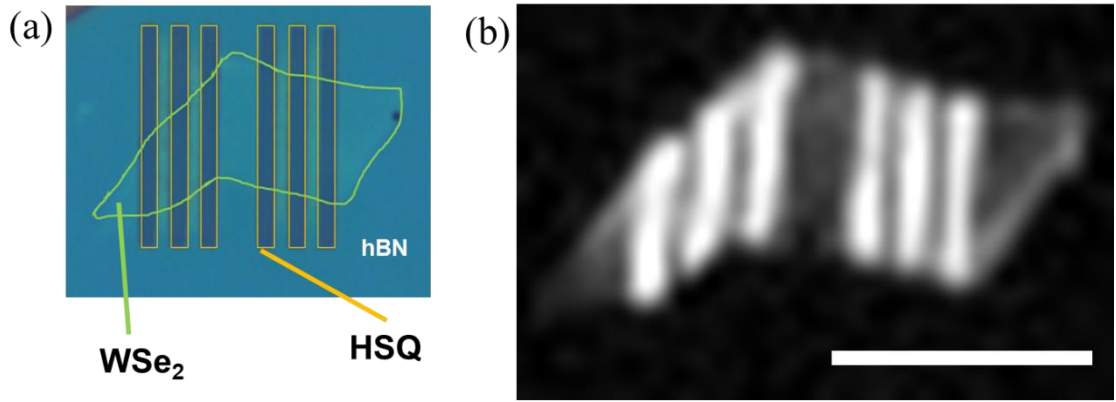

Figure 5 (a) An optical microscope image and (b) a corresponding Raman image of patterned Nb-doped $WSe_2$. Raman intensities from $A'_1$ and 2LA are used to make the Raman image. We used the Nb doping ratio larger than 10 % to enhance the difference in Raman intensity for this measurement. The scale bar corresponds 10 μm. The excitation wavelength of 532 nm, the excitation power of   μW, and exposure time of 30 seconds were used to obtain the Raman image. The Raman intensity is displayed by grayscale from 0 ~ 18000 counts.

**Electronic properties of Nb-doped $WSe_2$**

Figure 6a shows two-terminal transfer curves of $WSe_2$ before and after a 6-minutes exposure to a Nb beam; we used five monolayer $WSe_2$ grown on a $SiO_2$/Si substrate for this Nb doping. Ab-initio bandstructure calculation of Nb-doped $WSe_2$ (Fig. S4) shows that the Nb atom doped yields a localized state around VBM of $WSe_2$, showing that Nb should work as a p-dopant. As seen in transfer curves shown in Fig. 6a, on current ($I_{on}$) before doping is very small, typically ~ 100 pA, probably due to considerable contact resistance from the Schottky barrier at the interface between Ti and $WSe_2$. In contrast, after doping, $I_{on}$ is greatly enhanced in all cases, typically four orders in magnitude, and all devices clearly show p-type behavior consistent with previous works (Fig. 6a).(*24-26*) The observed $I_{on}$ increase should originate from the lowering of the Schottky barrier induced by the increase in hole density. The p-doping was also confirmed by an upward shift of Fermi energy in angle-resolved photoelectron spectra and Scanning tunneling microscopy/spectroscopy (Fig. S5 and S6). A critical point here is that p-type doping occurs uniformly all over the substrate because of the uniform intensity of Nb dopant beams; the current size of the Nb beam is at least 2 cm, which is larger than those of substrate used.

**DISCUSSION**

Because the present doping method is a post-doping method, it should be possible to increase the number of dopants afterward. Figure 6b shows a transfer curve of a pristine bilayer $WSe_2$ and transfer curves of the same bilayer $WSe_2$ after several consecutive Nb post-doping processes; electrodes can work even after multiple times Nb post-doping processes. As seen in Fig. 6a, all transfer curves show explicit p-type action, and $I_{on}$ increases in a stepwise manner as the doping process was repeated, finally reaching $10^{-6}$ A after three times doping; additional doping led to a decrease in $I_{on}$, probably due to increased Coulomb scatterings from increased impurities. The observed increase in $I_{on}$ demonstrates that we can tune the Nb doping ratio afterward, making the present post-doping method useful to fabricate various functional devices. One more critical point of the present post-doping method is that the same process should be transferred to dope various dopants to 2D semiconductors. We also performed Re doping with a Re beam to demonstrate this transferability and found that the post-doping method developed also works for Re doping (Fig. S7). PL spectra before and after Re doping show results similar to those obtained in the Nb doping, where the decrease in PL intensity is seen after Re doping. FET characteristic shows n-type action after the doping, which is expected in the case of Re doping.(*27*) Tuning doping afterward and the broad applicability make the present post-doping method useful to develop 2D-semiconductor-based nanoelectronics.

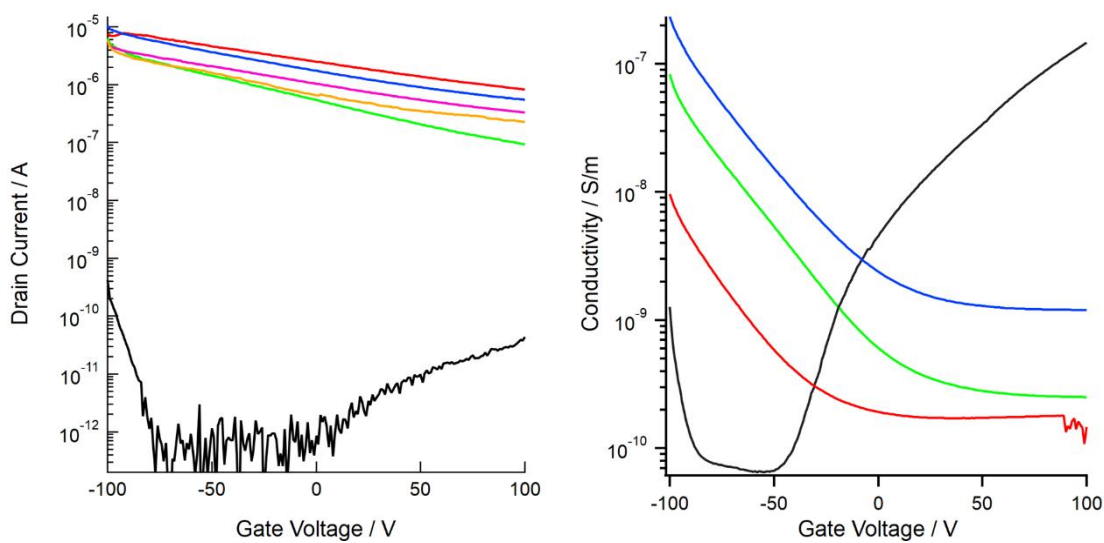

Figure 6 (a) Transfer curves of monolayer $WSe_2$ measured at room temperature. All devices are on the same substrate, and 6-minute Nb doping was performed before the measurements. The black curve represents a typical transfer curve of monolayer $WSe_2$ before the doping (channel

length = 20 μm, channel width = 20 μm).    (b) Transfer curves of pristine multilayer WSe$_2$ (black), WSe$_2$ after 1 min- (red), 2 min- (green), and 3 min-doping (blue). All transfer curves were measured at room temperature with a bias voltage of 10 V (channel length = 20 μm, channel width = 67μm).

**Methods**

Sample preparations

We prepared WSe$_2$ and MoSe$_2$ by the mechanical exfoliation or chemical vapor deposition (CVD) method on a SiO$_2$/Si substrate according to a previously reported procedure.(*28*) The prepared TMDs were put in a vacuum chamber (base pressure is 10$^{-6}$ Pa) and were heated up to 823 K. We started supplying Se (99.999 % purity, SigmaAldrich) with a Knudsen cell when the sample temperature becomes higher than 623 K to avoid unwanted formation of Se vacancies; supply rate of Se was kept constant (2 Å/sec.) throughout the doping process. This preheating process makes the sample surface clean, which is required for the following metal doping process. When the sample temperature becomes 523 K, we start supplying dopant (Nb or Re) with an electron beam evaporation (EFM 3, FOCUS GmbH) for 1~15 minutes with the supply rate of 0.67~1.11 monolayer/hour.

Characterizations

Raman and PL spectra were measured with microspectroscopy systems (Jovin-Yvon HR-800, and Renishaw InVia Raman) with 488 or 532 nm CW laser excitation. The laser was focused onto samples with objective lenses (×50–100 and 0.7–0.85 NA), and Raman and PL signals were detected with a charge-coupled device (CCD). All measurements were carried out at room temperature under atmospheric conditions. Electronic properties, including transfer curves, were measured with a semiconductor parameter analyzer (Kethley 4200-SCS/F) at room temperature. Optical responses were obtained with a homemade microspectroscopy system. HAADF- and ABF-STEM images were taken at room temperature by using a JEM-ARM200F ACCELARM (cold field emission gun) equipped with a CEOS ASCOR corrector, operated at 120 kV. For each frame of STEM image, fast scan rate of 3~ 5 μs per pixel is used. Several frames up to ten are overlapped after drift compensation to form a HAADF or ABF STEM image. EDS elemental mapping is performed by using JED-2300.

Ab-initio calculations

The geometric and electronic structure of Nb-doped $WSe_2$ were investigated using the STATE package based on density functional theory.(*29*) The generalized gradient approximation with the functional forms of Perdew-Burke-Ernzerhof is adopted to describe the exchange-correlation potential energy.(*30, 31*) The electron-ion interaction is treated by ultrasoft pseudopotentials.(*32*) Valence wave function and deficit charge density were expanded by plane-wave basis sets with the cutoff energies of 25 and 225 Ry, respectively. Ab initio molecular dynamics (MD) simulations were conducted with the use of the velocity scaling method to maintain the temperature at 300 K to implant Nb into $WSe_2$. Nb atom has initial velocities of 0.01, 0.3, and 1 eV toward the $WSe_2$ layer. Ab initio calculations were conducted with 2 x 2 and single Gamma points for static and MD calculations, respectively, under the 6x6 lateral periodicity of WSe2 with Nb atom. Atomic coordinates were optimized until the force acting on each atom became less than 5 mRy/A.

**Data availability**

The data that support the findings of this study are available from the corresponding author upon reasonable request.

formalism. *Physical Review B* **41**, 7892-7895 (1990).

Acknowledgements

R. K. was supported by JSPS KAKENHI Grant numbers JP16H03825, JP16H00963, JP15K13283, JP25107002, and JST CREST Grant Number JPMJCR16F3. T. H. was supported by JSPS KAKENHI Grant number JP19J15359. M. O. was supported by JSPS KAKENHI Grant number JP19K15403. K.W. and T.T. acknowledge support from the Elemental Strategy Initiative conducted by the MEXT, Japan, Grant Number JPMXP0112101001, JSPS KAKENHI Grant Numbers JP20H00354 and the CREST(JPMJCR15F3), JST.

## S1 Optical images of a monolayer WSe$_2$ before and after Nb doping

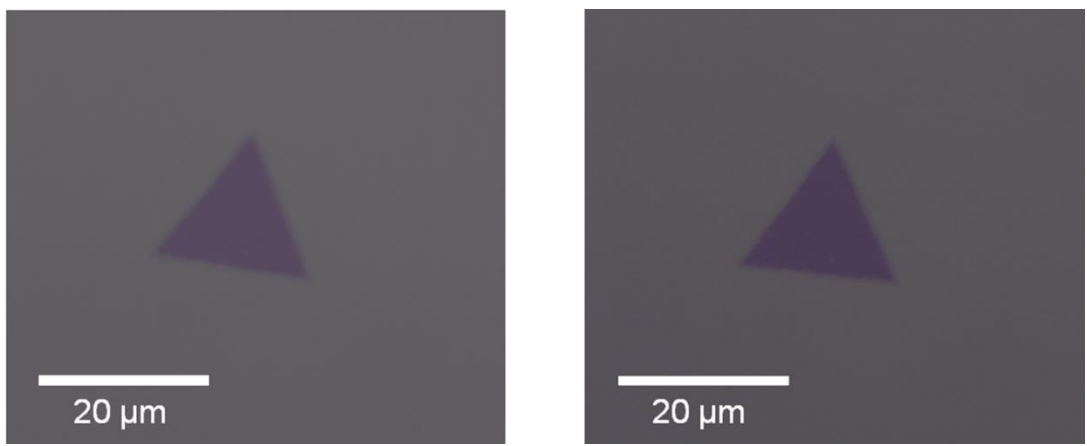

Fig. S1 A typical optical microscope image of a monolayer WSe$_2$ before (left) and after Nb doping (right).

In this case, Nb doping was performed toward a CVD-grown monolayer WSe$_2$ for 6 minutes, which corresponds to 7.4 % Nb doping. As seen in these images, the two images show almost no differences, demonstrating that the doping process does not cause the formation of cracks and wrinkles.

## S2 A typical EDS spectrum and the corresponding elemental mapping

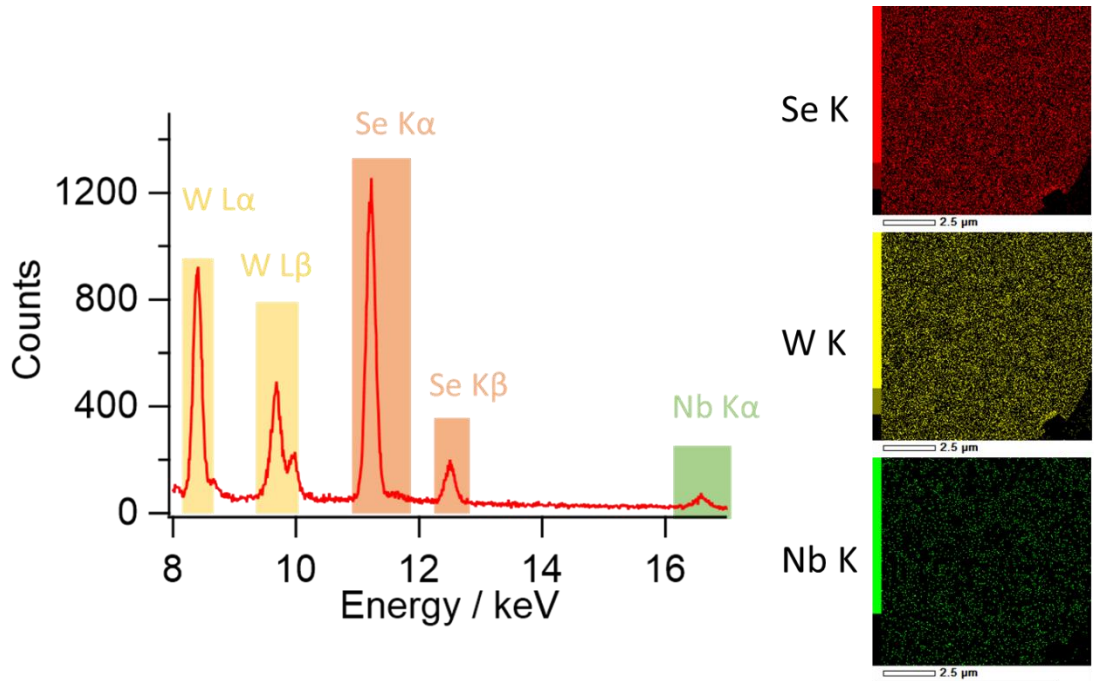

Fig. S2 A typical EDX spectrum and the corresponding elemental mapping of Se, W, and Nb.

EDX spectrum shown here shows uniform distribution of Se, W, and Nb over the observed area. This uniform distribution is consistent with transfer curves of FETs on a substrate, where all FETs at different locations show p-type action.

## S3 Snapshots of a Nb atom injected towards monolayer WSe$_2$

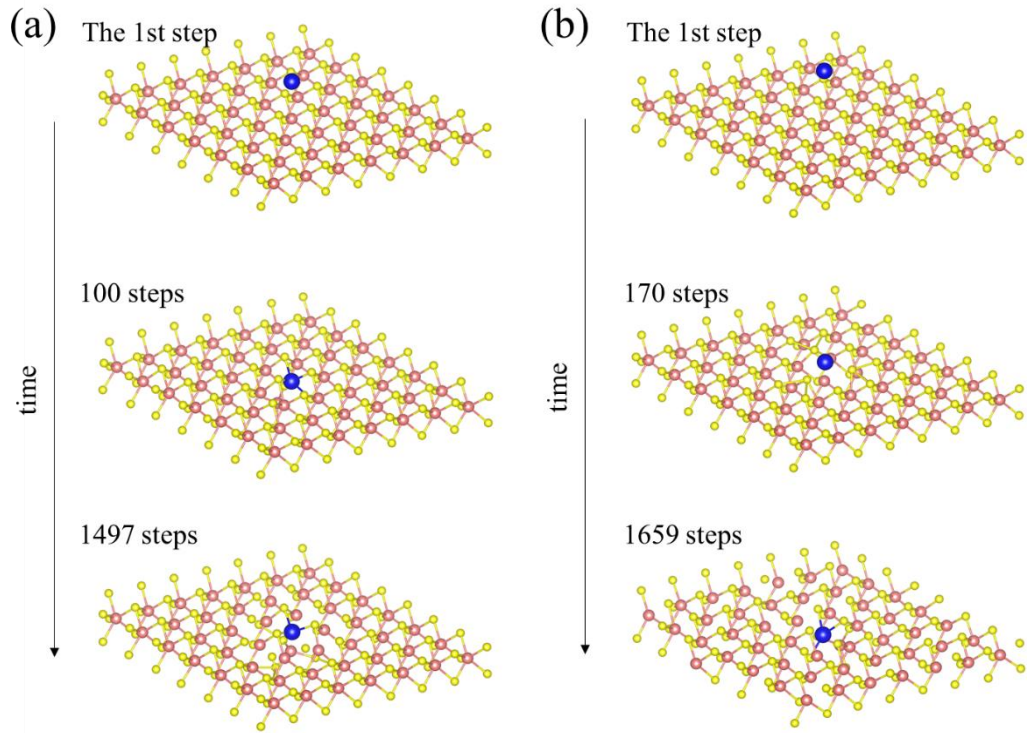

Fig.S3 Snapshots of a Nb atom injected toward a monolayer WSe$_2$. The kinetic energy ($E_{kin}$) of Nb is 10 meV (left) and 1 eV (right). Each step corresponds to 2 fs.

In addition to the case with $E_{kin}$ of 1 eV, a Nb atom is successfully incorporated into the framework of WSe$_2$, even in the case of $E_{kin}$ of 10 meV. In the actual experiment, we also supply Se atoms with high-flux value, and the reaction time scale is much longer than those of simulation (10 min scale in the experiments, whereas ca. 3 ps in simulation). Thus, annealing with excess Se atoms for a long time may lead to structural reconstruction to form the regulated hexagonal network of WSe$_2$.

## S4 Bandstructure of Nb-doped monolayer WSe₂ and corresponding wavefunctions

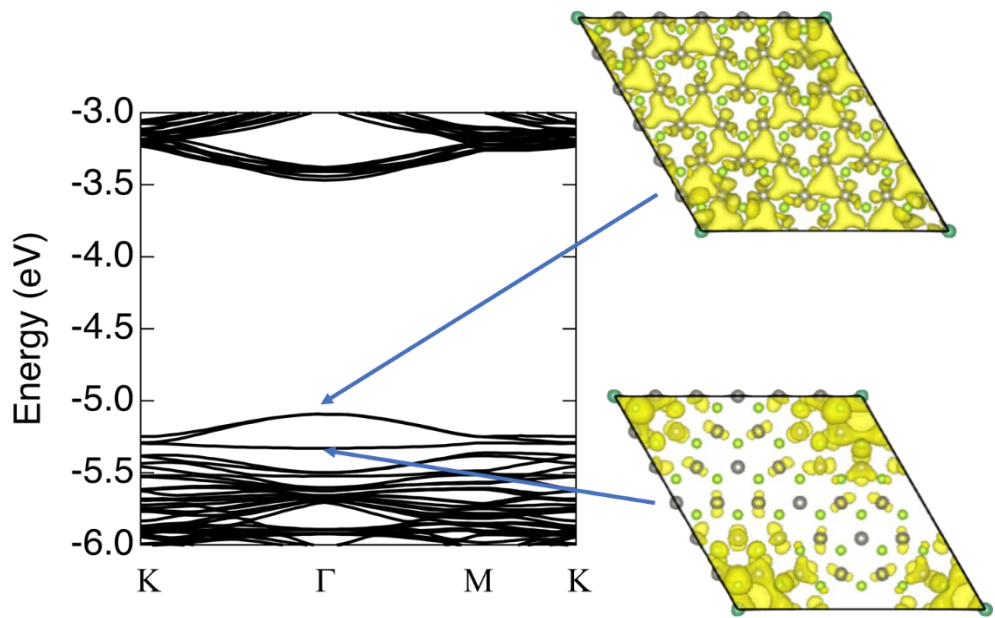

Fig. S4 Bandstructure of Nb-doped monolayer WSe$_2$ and corresponding wavefunctions

This calculation has been performed with a 5 x 5 supercell, and both the valence band top and the conduction band bottom locate at the Γ point in the Brillouin zone of this supercell. Nb atoms locate at the corners of the unit cell, and the wavefunction of the relatively flat band localize at the Nb atoms, being consistent with that Nb works as a p dopant.

## S5 Band dispersion of pristine trilayer MoSe$_2$ and Nb-doped trilayer MoSe$_2$

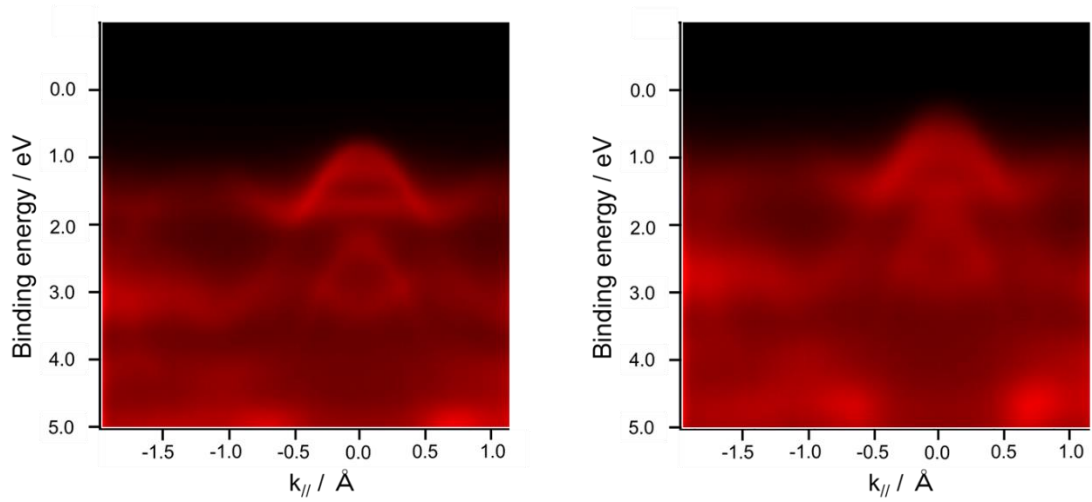

Fig. S5 Band dispersion of pristine MoSe$_2$ (right) and Nb-doped MoSe$_2$ (left) at room temperature.

As seen in the figures, band dispersion becomes broad after a Nb doping process. Although the broadening that probably arises from an increase of inhomogeneity exists, band dispersion is still clearly observable. Also, the valence band top of MoSe$_2$ after doping shifted to the higher energy side, which is consistent with the incorporation of p-dopant, Nb.

## S6 STM and STS observations of a Nb in a Nb-doped WSe$_2$

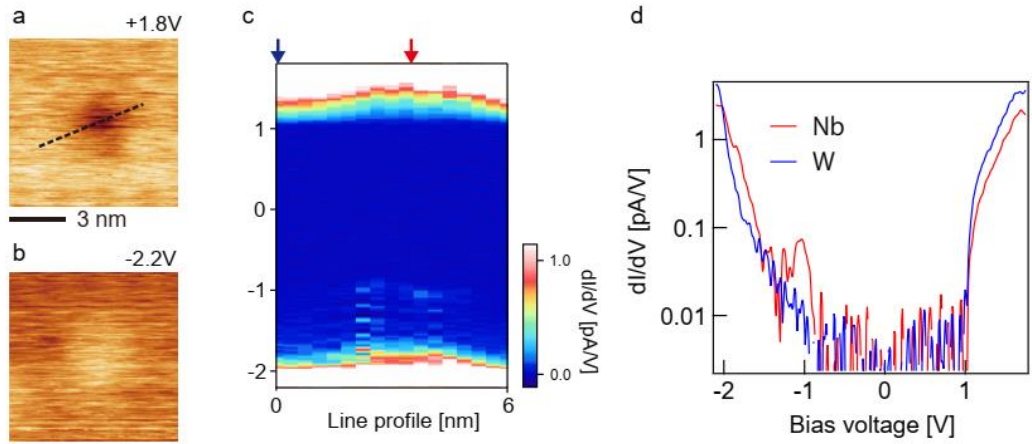

Fig. S6 (a) and (b) STM images of WSe$_2$ around a doped Nb taken at vias voltage of 1.8 and -2.2 V, respectively. (c) and (d) a corresponding STS mapping and an STS spectrum, respectively.

As shown in the STM images, a doped Nb gives a delocalized electronic structure, whose size is around 3 nm. The STS spectrum shows an upward shift in energy, indicating the existence of a positive charge around the doped Nb.

## S7 Electronic and optical measurements of Re-doped samples

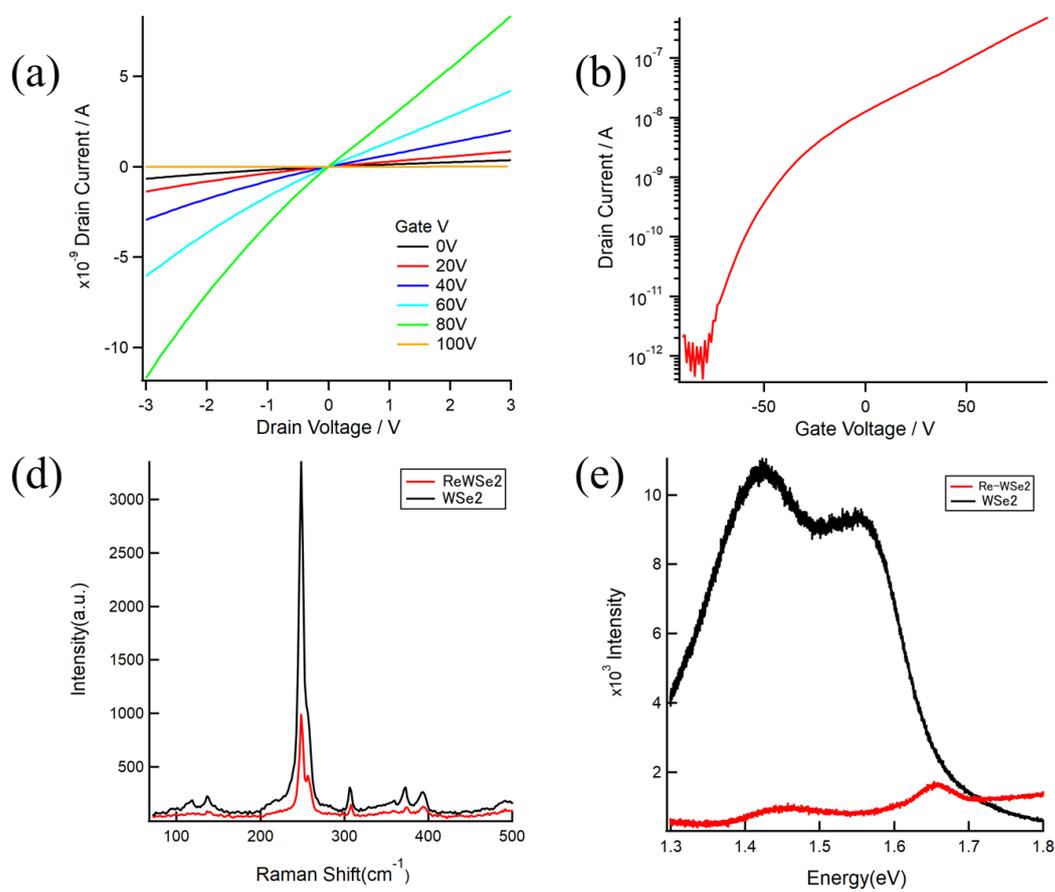

Fig. S6 (a) and (b) show output curves and a transfer curve of monolayer MoSe$_2$ measured at room temperature, respectively. (d) and (e) show Raman and PL spectra of trilayer WSe$_2$ before and after Re doping.